\documentstyle[12pt,a4]{article}
\begin{document}

\begin{center}
{\bf On The Baxter's Q - Operator for the XXX Spin Chain.}\\
\vskip \baselineskip
G.P.Pronko \\
{\it Institute for High Energy Physics, Protvino,Moscow reg. 142284,
Russia\\
International Solvay Institute, Brussels, Belgium.}
\end{center}

\vskip 1in
\begin{abstract} \normalsize
\par
We discuss the construction of Baxter's $Q$-operator. The suggested
approach leads to the one-parametric family of $Q$-operators, satisfying to
the wronslian-type relations. Also we have found the generalization of
Baxter operators, with defines the nondiagonal part of the monodromy.
\end{abstract}

\section {Introduction.}

Long ago, considering $XYZ$ spin chain Baxter\cite{Baxter} has introduced
so called $Q$- operator, which permitted him to find the eigenvalues of the
transfer matrix $t(x)$ in spite of the absence of Bethe Ansatz for this spin
chain. This object may be defined by the following operator equation:
\begin{equation} 
t(x)Q(x)=a(x)Q(x+i)+b(x)Q(x-i),
\end{equation}
together with the requirements
\begin{equation} 
[t(x),Q(y)]=[Q(x),Q(y)]=0.
\end{equation}
Recently, in the series of papers Bazhanov, Lukyanov and Zamolodchikov
\cite{BLZ} have given an explicit construction of such operators for the
case of certain integrable field model. Moreover their construction
definitely gives the pair the of operators $Q_{\pm}(x)$, satisfying apart
from (1) and (2) also
\begin{equation} 
[Q_{+}(x),Q_{-}(y)]=0
\end{equation}
and certain "wronskian" relation. which becomes the origin of the various
fusion relations.

However, the extension of their results for six-vertex  spin chain requires
an external magnetic field which cannot be eliminated by limiting
procedure. Therefore in the simplest case of $XXX$ spin chain we do not
know $Q_{\pm}(x)$ -operators, though from the point of view of quantum
inverse scattering method (QISM) \cite{Faddeev1} their construction should
be universal for any integrable system. In \cite{PS} we investigated the
equation (1) for the eigenvalues of the transfer matrix in the cases of $XXX$
and $XXZ$ and had proven that there exists a pair of solutions (we called
it $Q(x)$ and $P(x)$) which are the polynomials (or trigonometric
polynomials for $XXZ$ case) in spectral parameter and which also satisfies
"wronskian" relations.

In the present paper we give the explicit operator construction of the
one-parametric solution of the equation (1) and also the solutions of
generalized equation, where instead of the transfer matrix enters the whole
monodromy matrix.

To simplify the discussion we shall consider the case of quantum spin
$1/2$. The generalization for arbitrary quantum spin as well as for
inhomogeneous chain are more or less straightforward.

In the frameworks of QISM \cite{Faddeev1}, the monodromy matrix $T^{l}(x)$
is defined as the ordered product of $L$ - operators, acting in the $2\times
(2l+1)$ - dimensional space:
\begin{equation} 
L^{l}_{n}=x+2is^{a}_{n}L^{a},
\end{equation}
where $s^{a}_{n}$ are the operators of quantum spin $1/2$, while $L^{a}$
operators, acting in the auxiliary $2l+1$ - dimensional space. The monodromy
matrix is then given by
\begin{equation} 
T^{l}(x)=\prod_{n=1}^{N}L^{l}(x),
\end{equation}
where $N$- is the length of the chain
and the transfer matrix $t^{l}(x)$ is the trace of $T^{l}(x)$ over the
auxiliary space:
\begin{equation} 
t^{l}(x)=Tr T^{l}(x).
\end{equation}
Note that for the case isotropic $XXX$ - spin chain, the monodromy matrix
$T^{l}(x)$ is the scalar with respect to simultaneous rotation in the
quantum and auxiliary spaces
\begin{equation} 
[\vec S +\vec L, T^{l}(x)]=0,
\end{equation}
where
\begin{equation} 
\vec S=\sum_{N}^{n=1}\vec s_{n}.
\end{equation}
Therefore the transfer matrix $t^{l}(x)$ is the scalar with respect to
quantum spin:
\begin{equation} 
[\vec S, t^{l}(x)]=0.
\end{equation}
The full set of commutation relations between matrix elements of the
monodromy matrix with different spectral parameters is contained in the
following equation \cite{Faddeev1}:
\begin{equation} 
R^{l,l'}(x-y)T^{l}(x)T^{l'}(y)=T^{l'}(y)T^{l}(x)R^{l,l'}(x-y),
\end{equation}
where the monodromies $T^{l}(x)$ and $T^{l'}(y)$ have common quantum space
and different auxiliary spaces. The $R$-matrix, which acts in the tensor
product of auxiliary spaces with dimension $(2l+1)\times (2l'+1)$ is the
function of total auxiliary spin $\vec J=\vec L+\vec L'$ \cite{Faddeev1}:
\begin{equation} 
R^{l,l'}(x)=e^{i\pi J}\frac{\Gamma(J+1-ix)}{\Gamma(J+1+ix)},
\end{equation}
where the operator $J$ is given by:
\begin{equation} 
J=\left(\vec J^{2}+1/4\right)^{1/2}-1/2.
\end{equation}
The same commutation relations, as (10) are valid also for $L$-operators:
\begin{equation} 
R^{l,l'}(x-y)L^{l}(x)L^{l'}(y)=L^{l'}(y)L^{l}(x)R^{l,l'}(x-y),
\end{equation}
The equation (10) has many important corollaries, among which there are so
called fusion relations. The later plays the key role in what follows.
One of these fusion relation for transfer matrix has the following form:
\begin{equation} 
t^{1/2}(x)t^{l}(x+i(l+1/2))=(x+\frac{i}{2})^{N}t^{l+1/2}(x+il)+(x-
\frac{i}{2})^{N}t^{l-1/2}(x+i(l+1)).
\end{equation}
If we denote as $A(x,l)$ the transfer matrix with shifted spectral
parameter
\begin{equation} 
A(x,l)=t^{l}(x+i(l+1/2)),
\end{equation}
the relation (14) will take the form
\begin{equation} 
t^{1/2}(x)A(x,l)=(x+\frac{i}{2})^{N}A(x-i,l+1/2)+(x-\frac{i}{2})^{N}
A(x+i,l-1/2),
\end{equation}
which is very similar to the defining relation for Baxter $Q$- operator in
the case of $XXX$ spin chain \cite{PS}:
\begin{equation} 
t^{1/2}(x)Q(x)=(x+\frac{i}{2})^{N}Q(x-i)+(x-\frac{i}{2})^{N}Q(x+i).
\end{equation}
The difference of (16) and (17) is due to the shift of second argument of
$A(x,l)$ in the r.h.s. of (16). To eliminate this difference we shall make
the following trick. Let us forget for a moment that $l$ denotes the
representation of auxiliary spin, and takes only integer or half-integer
values and consider $A(x,l)$ as a function of two complex arguments. Then
the new function, which is defined as
\begin{equation} 
Q(x,l)=A(x,l+ix/2)
\end{equation}
apparently satisfies the relation
\begin{equation} 
t^{1/2}(x)Q(x,l)=(x+\frac{i}{2})^{N}Q(x-i,l)+(x-\frac{i}{2})^{N}Q(x+i,l).
\end{equation}

In such a way we obtain a one parametric family of operators, satisfying
Baxter equation (17). Of course, this construction can not be considered as
rigorous, because the analytic continuation from countable set of points
into complex plane is not unique and we have used this trick only to
illustrate the idea. In what follows we shall give more educated
construction of the $Q$-operator, based on this idea. However, if we
impose the condition, that after the continuation into complex $l$-plane,
the operator $A(x,l)$ remains polynomial in $l$, then this continuation
becomes unique and this trick gives the effective way for calculation of
eigenvalues of the $Q$-operators via the eigenvalues of the monodromies
$t^{l}(x)$.

\section {The $\cal L$-operators}

The discussion of the previous section made it clear that for the
construction of the $Q$-operator we need the complex spin in the
auxiliary space. Also, we shall look for the $Q$-operator in the form of
the trace of appropriate modoromy:
\begin{equation} 
Q(x)=Tr \hat Q(x),
\end{equation}
where the operator $\hat Q(x)$ acts in the tensor product of quantum space
of $s=1/2$ and infinite dimensional auxiliary space $\Gamma$. As we shall
see for our purpose it is sufficient that this space $\Gamma$ is the
representation of the algebra:
\begin{equation} 
[\rho_{i},\rho^{+}_{j}]=\delta_{ij},\qquad    i,j=1,2
\end{equation}
The operator $\hat Q(x)$ will be given by the ordered product:
\begin{equation} 
\hat Q(x)=\prod_{n=1}^{N}{\cal L}_{n}(x),
\end{equation}
where ${\cal L}_{n}(x)$-are the operators, depending on $\rho$ and
$\rho^{+}$ and acting in the space of $n$-th quantum spin.

Further we shall need to consider the product
$\left(L^{1/2}_{n}(x)\right)_{ij} {\cal L}_{n}(x)$, which acts in the
auxiliary space $\Gamma\times 2$ ($\Gamma$ - for ${\cal L}_{n}(x)$ and 2 -
is two-dimensional auxiliary space for $L^{1/2}_{n}(x)$). In this space it
is convenient to consider a pair of projectors $\Pi^{\pm}_{ij}$:
\begin{eqnarray} 
\Pi^{+}_{ij}&=&(\rho^{+}\rho
+1)^{-1}\rho_{i}\rho^{+}_{j}=\rho_{i}\rho^{+}_{j} (\rho^{+}\rho
+1)^{-1},\nonumber\\
\Pi^{-}_{ij}&=&(\rho^{+}\rho
+1)^{-1}\epsilon_{il}\rho^{+}_{l}\epsilon_{jm}
\rho_{m}=\epsilon_{il}\rho^{+}_{l}\epsilon_{jm}\rho_{m}(\rho^{+}\rho
+1)^{-1},
\end{eqnarray}
where
\begin{eqnarray} 
\rho^{+}\rho&=&\rho^{+}_{i}\rho_{i}\nonumber\\
\epsilon_{ij}&=&-\epsilon_{ji}, \quad \epsilon_{12}=1.
\end{eqnarray}
These projectors formally satisfy the following relations:
\begin{eqnarray} 
\Pi^{\pm}_{ik}\Pi^{\pm}_{kj}&=&\Pi^{\pm}_{ij},\nonumber\\
\Pi^{+}_{ik}\Pi^{-}_{kj}&=&0,\nonumber   \\
\Pi^{+}_{ij}+\Pi^{-}_{ij}&=&\delta_{ij}.
\end{eqnarray}
Rigorously speaking the r.h.s. of the first equation (25) in the Fock
representation has an extra term, proportional to the projector on the
vacuum state, but, as we shall see below, this term is irrelevant in the
present discussion.

In order to define $Q$ - operator which satisfies Baxter equation (17) we
shall exploit Baxter's idea \cite{Baxter}, which we reformulate as
following: ${\cal L}_{n}(x)$-{\it operator should satisfies the
relation}:
\begin{equation} 
\Pi^{-}_{ij}\left(L^{1/2}_{n}(x)\right)_{jl}{\cal L}_{n}(x)\Pi^{+}_{lk}=0.
\end{equation}
If this condition is fulfilled, then
\begin{eqnarray} 
\left(L^{1/2}_{n}(x)\right)_{ij}{\cal
L}_{n}(x)&=&\Pi^{+}_{ik}\left(L^{1/2}_{n}(x)\right)_{kl}{\cal
L}_{n}(x)\Pi^{+}_{lj} +\nonumber\\
\Pi^{-}_{ik}\left(L^{1/2}(x)_{n}\right)_{kl}{\cal
L}_{n}(x)\Pi^{-}_{lj}&+&\Pi^{+}_{ik}\left(L^{1/2}_{n}(x)\right)_{kl}{\cal
L}_{n}(x)\Pi^{-}_{lj}.
\end{eqnarray}
In other words, the condition (26) guaranties that the r.h.s. of (27) in
the sense of projectors $\Pi^{\pm}$ has the triangle form and this form
will be conserved for products over $n$ due to orthogonality of the
projectors.

From (26) we obtain
\begin{equation} 
\epsilon_{jm}\rho_{m}\left(x\delta_{jk}+is^{a}_{n}\sigma^{a}_{jk}\right)
{\cal L}_{n}(x)\rho_{k}=0,
\end{equation}
or
\begin{equation} 
{\cal L}_{n}(x)\rho_{k}=\left[is^{a}_{n}\sigma^{a}_{kl}-\delta_{kl}\right]
\rho_{l}A_{n}(x)
\end{equation}
and
\begin{equation} 
\epsilon_{jm}\rho_{m}\left[x\delta_{jk}-is^{a}_{n}\sigma^{a}_{jk}\right]
{\cal L}_{n}(x)=B_{n}(x)\epsilon_{kl}\rho_{l},
\end{equation}
where $A_{n}(x)$ and $B_{n}(x)$ are some operators which we shall find now.
Making use of (29) let us rewrite the first term in the r.h.s. of (27) in
the following form:
\begin{equation} 
\Pi^{+}_{ik}\left(L^{1/2}_{n}(x)\right)_{kl}{\cal L}_{n}(x)\Pi^{+}_{lj}=
-(x+i/2)(x-3i/2)\rho_{i}A_{n}(x)\rho^{+}_{j}(\rho^{+}\rho+1)^{-1}
\end{equation}
This equation may also be written as
\begin{equation} 
\Pi^{+}_{ik}\left(L^{1/2}_{n}(x)\right)_{kl}{\cal L}_{n}(x)\Pi^{+}_{lj}=
(x+i/2)\rho_{i}{\cal L}_{n}(x-i)\rho^{+}_{j}(\rho^{+}\rho+1)^{-1},
\end{equation}
provided the operator $A_{n}(x)$ is given by
\begin{equation} 
A_{n}(x)=-(x-3i/2)^{-1}{\cal L}_{n}(x).
\end{equation}
Substituting (33) into (29) we obtain the desired equation for ${\cal L}$
operator:
\begin{equation} 
\left(s^{a}_{n}\sigma^{a}_{ij}+ix\delta_{ij}\right)\rho_{j}{\cal L}_{n}(x)=
(1/2+ix){\cal L}_{n}(x+i)\rho_{i}.
\end{equation}
If this equation is satisfied, we immediately find the operator $B_{n}(x)$
in (30):
\begin{equation} 
B_{n}(x)=(x-i/2){\cal L}_{n}(x+i).
\end{equation}
Having (35) we can also rewrite the second term in the r.h.s. of (27), as
we did in the equation (32) and finally arrive at
\begin{eqnarray} 
&\left(L^{1/2}_{n}(x)\right)_{ij}{\cal L}_{n}(x)=
(x+i/2)\rho_{i}{\cal
L}_{n}(x-i)\rho^{+}_{j}(\rho^{+}\rho+1)^{-1}&\nonumber\\
&+(x-i/2)(\rho^{+}\rho+1)^{-1}\epsilon_{il}\rho^{+}_{l}{\cal
L}_{n}(x+i)\epsilon_{jm}\rho_{m}+
\Pi^{+}_{ik}\left(L^{1/2}_{n}(x)\right)_{kl}{\cal
L}_{n}(x)\Pi^{-}_{lj}&.
\end{eqnarray}
We do not care to rewrite the last term in the r.h.s. of (36) because this
term does not contribute into the final expression of $Q$-operator.

Until now our discussion  was quite formal because we did not specify the
representation of the algebra (21). The detailed investigation of the
equation (34) shows that the usual Fock representation for (21) does not
fit for our purpose, therefore we shall use less restrictive holomorphic
representation.

Let the operator $\rho^{+}_{i}$ be the operator of multiplication by the
$\alpha_{i}$, while the operator $\rho_{i}$-the operator of
differentiation with respect to $\alpha_{i}$:
\begin{eqnarray} 
\rho^{+}_{i}\psi(\alpha)&=&\alpha_{i}\psi(\alpha),\nonumber\\
\rho_{i}\psi(\alpha)&=&\frac{\partial}{\partial \alpha}\psi(\alpha).
\end{eqnarray}
The operators $\rho^{+}_{i},\rho_{i}$ are canonically conjugated for the
scalar product:
\begin{equation} 
(\psi,\phi)=\int \frac{\prod_{i=1,2}d\alpha_{i}d\bar\alpha_{i}}{(2\pi
i)^{2}} e^{-\alpha\bar\alpha}\bar\psi(\alpha)\phi(\alpha)
\end{equation}
The action of an operator in holomorphic representation is defined by its
kernel:
\begin{equation} 
\left(K\psi\right)(\alpha)=\int d\mu(\beta,\bar\beta)K(\alpha,\bar\beta)
\psi(\beta),
\end{equation}
where we have denoted
\begin{equation} 
d\mu(\beta,\bar\beta)=\frac{\prod_{i=1,2}d\alpha_{i}d\bar\alpha_{i}}{(2\pi
i)^{2}}.
\end{equation}
In this frameworks we can formulate the following

{\it Theorem} The kernel of the operator ${\cal L}_{n}(x)$ , satisfying
equation (34) in holomorphic representation has the following form:
\begin{equation} 
{\cal L}_{n}(x,l,\alpha,\bar\beta)=\left[x+\frac{i}{2}(\rho^{+}\rho+1)+
is^{a}_{n}\rho^{+}\sigma^{a}\rho\right]\frac{(\alpha\bar\beta)^{2l+ix}}
{\Gamma(2l+ix+1)},
\end{equation}
where $l$ is arbitrary number.

The proof is trivial by direct substitution of (41) into (34), using the
definition (39). In such a way, the operator ${\cal L}_{n}(x)$ given
by (41) solves the equation (36) for left multiplication by $L^{1/2}(x)$.
Changing the order of the multiplication in (36), we can prove that
\begin{eqnarray} 
&{\cal L}_{n}(x)\left(L^{1/2}_{n}(x)\right)_{ij}=
(x+i/2)(\rho^{+}\rho+1)^{-1}\rho_{i}{\cal
L}_{n}(x-i)\rho^{+}_{j}&\nonumber\\
&+(x-i/2)\epsilon_{il}\rho^{+}_{l}{\cal
L}_{n}(x+i)\epsilon_{jm}\rho_{m}(\rho^{+}\rho+1)^{-1}+
\Pi^{-}_{ik}\left(L^{1/2}_{n}(x)\right)_{kl}{\cal
L}_{n}(x)\Pi^{+}_{lj}&,
\end{eqnarray}
provided ${\cal L}_{n}(x)$ satisfies the equation:
\begin{equation} 
{\cal
L}_{n}(x)\rho^{+}_{i}\left(s^{a}_{n}\sigma^{a}_{ij}+ix\delta_{ij}\right)=
(1/2+ix)\rho^{+}_{j}{\cal L}_{n}(x+i).
\end{equation}
The direct substitution of (41) into (43) shows that (41) is the
solution also of this equation. The solution (41) possesses the invariance
with respect to simultaneous rotation in quantum and auxiliary spaces, as
an $L$-operator of $XXX$ chain:
\begin{equation} 
[\vec
s_{n}+\rho^{+}\frac{\vec\sigma}{2}\rho, {\cal L}_{n}(x)]=0.
\end{equation}

\section {The $Q$-operators}.

To proceed further we need to remind the definition of trace of an
operator in holomorphic representation. If the operator is given by its
kernel $F(\alpha,\bar \beta)$ then, (see e.g. \cite{Berezin})
\begin{equation} 
Tr F= \int d\mu(\alpha, \bar\alpha) F(\alpha,\bar\alpha),
\end{equation}
where the measure was defined in (40). Let us now consider the ordered
product of the ${\cal L}_{n}(x)$-operators, introduced in the previous
section
\begin{eqnarray} 
\hat Q(x,l,\alpha,\bar\beta)&=&\int
\prod_{i=1}^{N-1}d\mu(\gamma_{i},\bar\gamma_{i})
{\cal L}_{N}(x,l,\alpha,\bar\gamma_{N-1} )
{\cal L}_{N-1}(x,l,\gamma_{N-1},\bar\gamma_{N-2})\times\cdots\nonumber\\
\cdots&\times&{\cal L}_{2}(x,l,\gamma_{2},\bar\gamma_{1})
{\cal L}_{1}(x,l,\gamma_{1},\bar\beta).
\end{eqnarray}
Due to triangle (in the sense of projectors $\Pi^{\pm}$ ) structure of the
r.h.s. of (36) we obtain the following rule of multiplication of the
monodromy matrix $T^{1/2}(x)$ on operator $\hat Q$:
\begin{eqnarray} 
&\left(T^{1/2}(x)\right)_{ij}\hat Q(x,l,\alpha,\bar\beta)=
(x+\frac{i}{2})^{N}\rho_{i}\hat
Q(x-i,l,\alpha,\bar\beta)\rho^{+}_{j}(\rho^{+}\rho+1)^{-1}&\nonumber\\
&(x-\frac{i}{2})^{N}(\rho^{+}\rho+1)^{-1}\epsilon_{im}\rho^{+}_{m}\hat
Q(x+i,l,\alpha,\bar\beta)\epsilon_{jk}\rho_{k}+\Pi^{+}_{im}
\bigl(\cdots\bigr)_{mk}\Pi^{-}_{kj},&
\end{eqnarray}
where we omitted the explicit expression of the last term by obvious
reasons. In the derivation of (47) we have used the remnants of the
projectors $\Pi^{\pm}$  which govern the proper multiplication of each
term in (36) separately.

Now we can perform the trace operation over the holomorphic variables and
over $i,j$ indexes, corresponding to the auxiliary $2$-dimensional space
of $T^{1/2}(x)$. The result is the desired Baxter equation:
\begin{equation} 
t^{1/2}(x)Q(x,l)=(x+\frac{i}{2})^{N}Q(x-i,l)+(x-\frac{i}{2})^{N}Q(x+i,l),
\end{equation}
where, according to (45) and (46)
\begin{equation} 
Q(x,l)=\int d\mu(\alpha,\bar\alpha)\hat Q(x,l,\alpha,\bar\alpha).
\end{equation}
Note, that the trace of $\hat Q$ exists due to the exponential factor in
holomorphic measure (40) and has cyclic property, therefore $Q(x,l)$ is
invariant under cyclic permutation of the quantum spins. Further, due to
property (44) we easily obtain that $Q(x,l)$ is invariant with respect to
rotations of the total quantum spin:
\begin{equation} 
[\vec S,Q(x,l)]=0,
\end{equation}
where $\vec S$ is given in (8).

Recall, that ${\cal L}_{n}(x)$-operators satisfy also the relation (42)
for the right multiplication by $L^{1/2}(x)$, therefore
\begin{equation} 
[t(x), Q(x,l)]=0.
\end{equation}

The expression (46) for the $\hat Q$-operator could be essentially
simplified. For that let us rewrite the equation (41) in the following
form:
\begin{equation} 
{\cal L}_{n}(x,l,\alpha,\bar\alpha)=
K_{n}(x)\frac{(\alpha\bar\beta)^{2l+ix}}{\Gamma(2l+ix+1)},
\end{equation}
where we have denoted via $K_{n}(x)$ the following operator:
\begin{equation} 
K_{n}(x)=x+\frac{i}{2}(\rho^{+}\rho +1)+is^{a}_{n}\rho^{+}\sigma^{a}\rho.
\end{equation}
Here $\rho,\rho_{+}$-are the operators, acting in (52) on the variable
$\alpha$. The action of the operator ${\cal L}_{n}(x)$ in the form (52) on
the function $\psi$, according to (39) has the following form:
\begin{equation} 
\left({\cal L}_{n}(x)\psi\right)(\alpha)=\int d\mu(\beta,\bar\beta)
K_{n}(x)\frac{(\alpha\bar\beta)^{2l+ix}}{\Gamma(2l+ix+1)}\psi(\beta).
\end{equation}
Using property of the measure $d\mu (\beta,\bar\beta)$ we can transfer the
action of the operator $K_{n}(x)$ from the variable $\alpha_{i}$ onto the
variable $\beta_{i}$ and rewrite (54) in the form:
\begin{equation} 
\left({\cal L}_{n}(x)\psi\right)(\alpha)=\int d\mu(\beta,\bar\beta)
\frac{(\alpha\bar\beta)^{2l+ix}}{\Gamma(2l+ix+1)}K_{n}(x)\psi(\beta).
\end{equation}
Proceeding this way in the representation (46) we can collect all operators
$K_{n}(x)$ in one place:
\begin{eqnarray} 
\hat Q(x,l,\alpha,\bar \beta)&=&\int
\prod_{i=1}^{N-1}d\mu(\gamma_{i},\bar\gamma_{i})
\frac{(\alpha\bar\gamma_{N-1})^{2l+ix}(\gamma_{N-1}\bar\gamma_{N-2})^{2l+ix}
\cdots(\gamma_{2}\bar\gamma_{1})^{2l+ix}}{\left[\Gamma(2l+ix+1\right]^{N-1}}
\nonumber\\
&\times&
\prod_{m=1}^{N}K_{m}(x)\frac{(\gamma_{1}\bar\beta)^{2l+ix}}{\Gamma(2l+ix+1)}.
\end{eqnarray}
Now all operators $K_{m}(x)$ act on the variable $\gamma_{1}$, and we can
perform integration over $\gamma_{k},\bar\gamma_{k}$ with $k=2,\cdots,N-1$.
This integration could be done with the help of the following formula:
\begin{equation} 
\int
d\mu(\gamma,\bar\gamma)\frac{(\alpha\bar\gamma)^{2l+ix}
(\gamma\bar\beta)^{2l+ix}}{\left[\Gamma(2l+ix+1)\right]^{2}}=
\frac{(\alpha\bar\beta)^{2l+ix}}{\Gamma(2l+ix+1)}.
\end{equation}
From (57) follows that the factor
$(\alpha\bar\beta)^{u}/\Gamma(u+1)$ is actually the kernel
of some projector. Using this property, we arrive at the following
representation for $\hat Q$:
\begin{equation} 
\hat Q(x,l,\alpha,\bar\beta)=\int d\mu(\gamma,\bar\gamma)
\frac{(\alpha\bar\gamma)^{2l+ix}}{\Gamma(2l+ix+1)}\prod_{m=1}^{N}
K_{m}(x)\frac{(\gamma\bar\beta)^{2l+ix}}{\Gamma(2l+ix+1)}.
\end{equation}
Due to the fact that in (58) the ordered product of $K_{n}(x)$ acts only on
the variable $\gamma_{i}$ , we can derive the expression for trace of $\hat
Q(x,l,\alpha,\bar\beta)$, performing one more integration:
\begin{equation} 
Q(x,l)=\int d\mu(\gamma,\bar\gamma)\prod_{m=1}^{N}K_{m}(x)
\frac{(\gamma\bar\gamma)^{2l+ix}}{\Gamma(2l+ix+1)}
\end{equation}

Needless to say that for integer or half-integer $l$ and $x=0$ the
expression $(\alpha\bar\beta)^{2l}/\Gamma(2l+1)$ coincides with the
kernel of the projector on the representation $l$ of $su(2)$, so that the
equation (59)is actually the desired prescription for analytic continuation
into complex momentum, naively suggested in the  Introduction.

\section {The Intertwining Relations.}

In this section we shall derive several intertwining relations for the
operators ${\cal L}_{n}(x)$ and $L^{1/2}(x)$ which permit us to prove the
commutativity of $Q(x,l)$ and some other important corollaries.
First of all let us consider the representation (52) for ${\cal L}_{n}(x)$
-operator. The formal operator $K_{n}(x)$, which enters into this
representation is nothing else, but usual $L_{n}(x)$-operator of $XXX$
-spin chain with infinite dimensional auxiliary space, with shifted
spectral parameter. The shift commutes with $K_{n}(x)$, so we can prove
pure algebraically the $R$-matrix form of commutation relation for
$K_{n}(x)$:
\begin{eqnarray} 
\lefteqn{
R^{\rho,\tau}\left(x-y+\frac{i}{2}(\rho^{+}\rho-\tau^{+}\tau)\right)
K^{\rho}_{n}(x) K^{\tau}_{n}(y)=}\nonumber\\
& &=K^{\tau}_{n}(y)K^{\rho}_{n}(x)
R^{\rho,\tau}\left(x-y+\frac{i}{2}(\rho^{+}\rho-\tau^{+}\tau)\right) ,
\end{eqnarray}
where $R^{\rho,\tau}(x)$ is given by equation (11) with
\begin{equation} 
\vec J=\rho^{+}\frac{\vec\sigma}{2}\rho+\tau^{+}\frac{\vec\sigma}{2}\tau.
\end{equation}
The indexes $\rho$ and $\tau$ at the operators $K_{n}$ and $R$
indicate different operators, acting in their auxiliary spaces.
For the products of ${\cal L}$-operators equation (60) implies
\begin{eqnarray} 
\lefteqn{R^{\rho,\tau}\left(x-y+\frac{i}{2}(\rho^{+}\rho-\tau^{+}\tau)\right)
K^{\rho}_{n}(x)
K^{\tau}_{n}(y)\frac{(\alpha\bar\beta)^{2l+ix}}{\Gamma(2l+ix+1)}
\frac{(\gamma\bar\delta)^{2m+iy}}{\Gamma(2m+iy+1)} }\nonumber\\
&&=K^{\tau}_{n}(y)K^{\rho}_{n}(x)
R^{\rho,\tau}\left(x-y+\frac{i}{2}(\rho^{+}\rho-\tau^{+}\tau)\right)
\frac{(\alpha\bar\beta)^{2l+ix}}{\Gamma(2l+ix+1)}
\frac{(\gamma\bar\delta)^{2m+iy}}{\Gamma(2m+iy+1)}\nonumber\\
&&=K^{\tau}_{n}(y)K^{\rho}_{n}(x)
\frac{(\alpha\bar\beta)^{2l+ix}}{\Gamma(2l+ix+1)}
\frac{(\gamma\bar\delta)^{2m+iy}}{\Gamma(2m+iy+1)}
R^{\rho,\tau}\left(x-y+\frac{i}{2}(\rho^{+}\rho-\tau^{+}\tau)\right)
\nonumber\\
&&
\end{eqnarray}
Few comments should be made to this equations. The holomorphic variables,
corresponding to the pair of operators $\rho^{+},\rho$ are $\alpha, \bar
\beta$ , to the pair $\tau^{+},\tau$ --- $\gamma,\bar\delta$. The equations
(62) should be understood as the short version of the long story with
integrals over the holomorphic variables with the functions, depending upon
$\beta, \delta$. The last step of the chain of the equations is due to the
same property of measure, which permitted the transition from (54) to (55).

Coming back to the ${\cal L}$ operators we can write the following
intertwining relation:
\begin{eqnarray} 
\lefteqn{ R^{\rho,\tau}(x-y+\frac{i}{2}(\rho^{+}\rho-\tau^{+}\tau)){\cal
L}_{n}(x,l,\alpha,\bar\beta){\cal
L}_{n}(y,m,\gamma,\bar\delta)=}\nonumber\\ &&{\cal
L}_{n}(y,m,\gamma,\bar\delta){\cal L}_{n}(x,l,\alpha,\bar\beta)
R^{\rho,\tau}(x-y+\frac{i}{2}(\rho^{+}\rho-\tau^{+}\tau))
\end{eqnarray}
Apparently, the same relation holds true also for $\hat Q$- operators. From
that we immediately obtain the commutativity of its traces:
\begin{equation} 
[Q(x,l),Q(y,m)]=0
\end{equation}

Further, let us consider another commutation relation \cite{Faddeev2} :
\begin{eqnarray} 
\lefteqn{ (x+\frac{i}{2}\vec\sigma_{1}\vec\sigma_{2})(y+i\vec\sigma_{1}\vec
M)(y-x+i\vec\sigma_{2}\vec M)}\nonumber\\
&&=(y-x+i\vec\sigma_{2}\vec M)(y+i\vec\sigma_{1}\vec M)
(x+\frac{i}{2}\vec\sigma_{1}\vec\sigma_{2}) ,
\end{eqnarray}
where the Pauly matrices $\sigma^{a}_{1},\sigma^{a}_{2}$ acts in their
spaces and $M^{a}$ are some operators satisfying
\begin{equation} 
[M^{a},M^{b}]=i\epsilon_{abc}M^{c}.
\end{equation}
In particular, we can set
\begin{equation} 
M^{a}=\rho^{+}\frac{\vec\sigma}{2}\rho,
\end{equation}
where $\rho^{+},\rho$ the Heisenberg variables (21). Now let us shift the
spectral parameter $y$ in (65) by $i/2(\rho^{+}\rho+1)$ and rewrite (65) in
the following form:
\begin{eqnarray} 
&\left(x+i\vec\sigma\vec
s_{n}\right)\left(y+\frac{i}{2}(\rho^{+}\rho+1)+i\vec
s_{n}\rho^{+}\vec\sigma\rho\right)
\left(y-x+\frac{i}{2}(\rho^{+}\rho+1)+i\vec\sigma\rho^{+}\frac{\vec\sigma}{2}
\rho\right)&\nonumber\\
&=\left(y-x+\frac{i}{2}(\rho^{+}\rho+1)+i\vec\sigma\rho^{+}\frac{\vec\sigma}
{2}\rho\right)\left(y+\frac{i}{2}(\rho^{+}\rho+1)+i\vec
s_{n}\rho^{+}\vec\sigma\rho\right)\left(x+i\vec\sigma\vec s_{n}\right),&
\nonumber\\
&&
\end{eqnarray}
where we interpret $\vec\sigma_{1}$ as quantum spin $\vec s_{n}$
while the $\vec\sigma_{2}$ serves as auxiliary spin. Equation (68) could be
also written as
\begin{equation} 
\left(L^{1/2}_{n}(x)\right)_{ik}K_{n}(y)\left(K(y-x+\frac{i}{2})\right)_{kj}
=
\left(K(y-x+\frac{i}{2})\right)_{ik}K_{n}(y)\left(L^{1/2}_{n}(x)\right)_{kj},
\end{equation}
where we explicitly wrote the indexes, corresponding to the auxiliary
space, index $n$ indicates corresponding quantum space and the operator
$K(x)$ was introduced in (53).

The equation (69) could be used for derivation of the intertwining relation
for $L^{1/2}$ and ${\cal L}$  operators. Indeed, let use consider the
following product of the operators, acting of the function in  holomorphic
representation
\begin{eqnarray} 
\lefteqn{\Biggl(\left(L^{1/2}_{n}(x)\right)_{ik}{\cal
L}_{n}(y,l)K_{kj}(y-x+\frac{i}{2})\psi\Biggr)(\alpha)}\nonumber\\
&&=\left(L^{1/2}_{n}(x)\right)_{ik}\int d\mu(\beta,\bar\beta)K_{n}(y)
\frac{(\alpha\bar\beta)^{2l+iy}}{\Gamma(2l+iy+1)}K_{kj}(y-x+\frac{i}{2})
\psi(\beta).
\end{eqnarray}
Moving the operator $K_{kj}(y-x+i/2)$ to the left, through the projector
$(\alpha\bar\beta)^{2l+iy}/\Gamma(2l+iy+1)$ and using (69) we obtain the
following relation:
\begin{equation} 
\left(L^{1/2}_{n}(x)\right)_{ik}{\cal L}_{n}(y,l)K_{kj}(y-x+\frac{i}{2})=
K_{ik}(y-x+\frac{i}{2}){\cal L}_{n}(y,l)\left(L^{1/2}_{n}(x)\right)_{kj},
\end{equation}
where the operator
\begin{equation} 
K_{ij}(x)=\left(x+\frac{i}{2}(\rho^{+}\rho+1)\right)\delta_{ij}+i\vec
\sigma_{ij}\rho^{+}\frac{\vec\sigma}{2}\rho
\end{equation}
plays the role of R-matrix.

The relation (71) gives rise to the analogous relation for the monodromies:
\begin{equation} 
\left(T^{1/2}(x)\right)_{ik}\hat
Q(y,l,\alpha,\bar\beta)K_{kj}(y-x+\frac{i}{2})
=K_{ik}(y-x+\frac{i}{2})\hat
Q(y,l,\alpha,\bar\beta)\left(T^{1/2}(x)\right)_{kj},
\end{equation}
from where we obtain the commutativity of the transfer matrix and our $Q$-
operator:
\begin{equation} 
[t(x),Q(y,l)]=0.
\end{equation}

\section {Again $Q$-Operators}

Now we are ready to discuss some important properties of Baxter's $Q$-
operator and it generalizations. Let us start from the Baxter equation (48)
for the $Q$-operator. Due to mutual commutativity of $Q(x,l)$ with
different spectral parameters and second arguments, we easily derive that
\begin{eqnarray} 
\lefteqn{
\left[\left(x-\frac{i}{2}\right)^{N}Q(x-i,l)-\left(x+\frac{i}{2}\right)^{N}
Q(x+i,l)\right]Q(x,m)}\nonumber\\
&&=\left[\left(x-\frac{i}{2}\right)^{N}Q(x-i,m)-\left(x+\frac{i}{2}\right)
^{N}Q(x+i,m)\right]Q(x,l),
\end{eqnarray}
or
\begin{equation} 
Q(x+\frac{i}{2},l)Q(x-\frac{i}{2},m)-Q(x-\frac{i}{2},l)Q(x+\frac{i}{2},m)
=C(l,m)x^{N},
\end{equation}
where $C(l,m)$ is some unknown operator, commuting with $Q$. To find
$C(l,m)$ we must calculate the l.h.s. of (76) for some convenient values of
arguments. From (59) follows that the $Q$- operator is proportional to the
trace of the projector, whose kernel in the holomorphic representation is
$(\gamma\bar\gamma)^{2l+ix}/\Gamma(2l+ix+1)$. This trace is given by:
\begin{equation} 
\int
d\mu(\gamma\bar\gamma)\frac{(\gamma\bar\gamma)^{2l+ix}}{\Gamma(2l+ix+1)}=
2(2l+ix+1).
\end{equation}
Note , that for $x=0$, the trace is $2\times$ dimension of the
representation of spin $l$. from (77) we conclude that
\begin{equation} 
Q(x,l)|_{ix=-(2l+1)}=0.
\end{equation}
Now let us set $x=i(2l+1/2)$ in the equation (76) (for $m,l$- integer or
half-integer and $m\geq l+1/2$). Then, due to (78) the first term in l.h.s.
of (76) disappears and we obtain:
\begin{equation} 
-Q(2il,l)Q(i(2l+1),m)=C(l,m)[i(2l+1)]^{N}
\end{equation}
Further, from (59) we derive that
\begin{equation} 
Q(2il,l)=t^{0}(i(2l+1))=[i(2l+1)]^{N}
\end{equation}
and
\begin{equation} 
Q(i(2l+1))=t^{m-l-1/2}(i(l+m)).
\end{equation}
Hence, the unknown coefficient $C(l,m)$ is given by
\begin{equation} 
C(l,m)=t^{m-l-1/2}(i(l+m)).
\end{equation}
For the case $l\geq m+1/2$ , we should put $x=i(2m+1/2)$ and $l$ and $m$
will change their places in the final answer. So, finally we obtain the
quantum wronskian in the following form:
\begin{equation} 
Q(x+\frac{i}{2},l)Q(x-\frac{i}{2},m)-Q(x-\frac{i}{2},l)Q(x+\frac{i}{2},m)
=-x^{N}t^{m-l-1/2}(i(l+m)).
\end{equation}
Note, that for $l=m$, the r.h.s. of (83) vanishes, as it should be for
wronskian of a linearly dependent solutions.

Proceeding along this way we can obtain the general relation involving the
transfer matrix with arbitrary spin in the auxiliary space in the r.h.s. of
(83) (the $x^{N}$is just the $t^{o}(x)$) (see \cite{BLZ},\cite{PS}). We
postpone the discussion of these relation for the future publication, where
we intend to give another its derivation .

Further we want to consider the generalization of Baxter equation, which
follows from the fundamental relation (47). To obtain these new relations,
let us multiply both sides of (47) by total spin in the auxiliary space:
\begin{eqnarray} 
\lefteqn{\left(\frac{1}{2}\sigma^{a}_{ik}+J^{a}\delta_{ik}\right)
\left(T^{1/2}(x)\right)_{ij}\hat Q(x,l,\alpha,\bar\beta)} \nonumber\\
&&=\left(\frac{1}{2}\sigma^{a}_{ik}+J^{a}\delta_{ik}\right)\Biggl
[(x+\frac{i}{2})^{N}\rho_{k}\hat Q(x-i,l,\alpha,\bar\beta)\rho^{+}_{j}
(\rho^{+}\rho+1)^{-1}\nonumber\\
&&+(x-\frac{i}{2})^{N}(\rho^{+}\rho+1)^{-1}\epsilon_{km}\rho^{+}_{m}\hat
Q(x+i,l,\alpha,\bar\beta)\epsilon_{jn}\rho_{n}\nonumber\\
&&+\Pi^{+}_{km}\bigl(\cdots\bigr)_{mn}\Pi^{-}_{nj}\Biggr],
\end{eqnarray}
where
\begin{equation} 
J^{a}=\rho^{+}\frac{\sigma^{a}}{2}\rho.
\end{equation}
Due to the obvious relations
\begin{eqnarray} 
\frac{1}{2}\sigma^{a}_{ik}\rho_{k}+J^{a}\rho_{i}&=&\rho_{i}J^{a},\nonumber\\
\frac{1}{2}\sigma^{a}_{ik}\epsilon_{km}\rho^{+}_{m}+J^{a}\epsilon_{ik}
\rho^{+}_{k}&=&\epsilon_{ik}\rho^{+}_{k}J^{a},\nonumber\\
\left(\frac{1}{2}\sigma^{a}_{ik}+J^{a}\delta_{ik}\right)\Pi^{pm}_{kj}&=&
\Pi^{pm}_{ik}\left(\frac{1}{2}\sigma^{a}_{kj}+J^{a}\delta_{kj}\right),
\end{eqnarray}
the equation (84) could be rewritten in the following form:
\begin{eqnarray} 
\lefteqn{\left(\frac{1}{2}\sigma^{a}_{ik}+J^{a}\delta_{ik}\right)\left(
T^{1/2}(x)\right)_
{kj}\hat Q(x,l,\alpha,\bar\beta)}\nonumber\\
&&=(x+\frac{i}{2})^{N}\rho_{i}J^{a}\hat
Q(x-i,l,\alpha,\bar\beta)\rho^{+}_{j}(\rho^{+}\rho+1)^{-1}\nonumber\\
&&+(x-\frac{i}{2})^{N}(\rho^{+}\rho+1)^{-1}\epsilon_{im}\rho^{+}_{m}J^{a}
\hat Q(x+i,l,\alpha,\bar\beta)\epsilon_{jk}\rho_{k}\nonumber\\
&&+\Pi^{+}_{ik}\left(\frac{1}{2}\sigma^{a}_{km}+J^{a}\delta_{km}\right)
\bigl(\cdots\bigr)_{ml}\Pi^{-}_{lj}.
\end{eqnarray}
If we now shall calculate the trace over the whole auxiliary space, the
last term in the r.h.s. of (87)  again will not contribute, as it was in
the case of usual Baxter equation and we obtain the following relation:
\begin{equation} 
\vec t^{1/2}(x)Q(x,l)+t^{1/2}(x)\vec Q(x,l)=(x+\frac{i}{2})^{N}\vec
Q(x-i,l)+ (x-\frac{i}{2})^{N}\vec Q(x+i,l),
\end{equation} where we have
introduced the notations:
\begin{eqnarray} 
\vec t^{1/2}(x)&=&Tr\frac{\vec\sigma}{2}T^{1/2}(x)\nonumber\\
\vec Q(x,l)&=& \int d\mu(\alpha,\bar\alpha)\vec J
\hat Q(x,l,\alpha,\bar\alpha)
\end{eqnarray}
This equation may be considered as inhomogeneous Baxter equation, where the
first term in the l.h.s. plays the role of inhomogeneity. Remarkably, that
the r.h.s. of (89) will not change if we simultaneously will change the
order of multiplication in both terms in the l.h.s.:
\begin{equation} 
Q(x,l)\vec t^{1/2}(x)+\vec Q(x,l)t^{1/2}(x)=(x+\frac{i}{2})^{N}\vec
Q(x-i,l)+ (x-\frac{i}{2})^{N}\vec Q(x+i,l).
\end{equation}
This property
could by derived either from equation (47), repeating all steps, which lead
us to (88) or directly from intertwining relation (73).  These new vector
$Q$-operators inherit many properties of the original Baxter operator. In
particularly, they also satisfy the wronskian-type relations, similar to
(83). We intend to present the detailed discussion of these operators in
separate publication. It worth to mention that we can go further,
multiplying the equation (47) by products of the generators of total
auxiliary spin. The relations (86) and triangle structure of the r.h.s. of
(47) guaranties that the last term will not contribute and we shall obtain
the relations, similar to the (90) for new tensorial generalizations of the
$Q$-operator. Also, multiplying both sides of (47) by the  operator
\begin{equation} 
U(\vec H)=\exp i\vec H(\frac{\vec\sigma}{2}+\vec J),
\end{equation}
where $\vec H$ are c-number, we shall obtain Baxter's operator for
$XXX$ spin chain in the external magnetic field.

\section {Concluding remarks.}

The construction of Baxter's $Q$ operator, considered in the present paper
for the case of $XXX$ spin chain seems to be rather universal and could be
extended for the case of anisotropic $XXZ$ spin chain. The key for this
generalization is again the "naive" analytic continuation suggested in the
Introduction.

Indeed, in the Baxter's parameterization \cite{Baxter}, the fusion
relation for $XXZ$ spin chain has the following form:
\begin{eqnarray} 
\lefteqn {t^{1/2}(\phi)t^{l}\left(\phi+(2l+1)\eta\right)}\nonumber\\
&&=\sin^{N}(\phi+\eta)t^{l+1/2}(\phi+2l\eta)
+\sin^{N}(\phi-\eta)t^{l-1/2}(\phi+2(l+1)\eta),
\end{eqnarray}
where $\eta$ is the crossing parameter. From (92) it is clear that the
function $Q(\phi,l)$, defined by
\begin{equation} 
Q(\phi,l)=t^{l-\phi/4\eta}(\phi/2+(2l+1)\eta)
\end{equation}
satisfies the relation:
\begin{equation} 
t^{1/2}(\phi)Q(\phi,l)=\sin^{N}(\phi+\eta)Q(\phi-2\eta,l)+
\sin^{N}(\phi-\eta)Q(\phi+2\eta,l).
\end{equation}
Again this trick can not be considered as the construction of the $Q$ -
operator, but it gives the strong evidence that the procedure described in
the present paper may be extended for the 6-vertex model. Very interesting
question is the further generalization of this approach to the case of
8-vertex spin chain, for which there exists also Baxter construction
\cite{Baxter} and for the case of the field model considered in \cite{BLZ}.

\section {Acknowledgments}

The author is grateful to V.Bazhanov, L.Faddeev, S.Sergeev, Yu.Stroganov,
A.  Volkov for their interest, discussions, criticism and encouragement.
This work was supported in part by ESPIRIT project NTCONGS and RFFI Grant
98-01-0070.

\end{document}